\documentclass{aa}  

\usepackage[normalem]{ulem}
\usepackage{graphicx}
\usepackage{color}
\usepackage{url}
\usepackage[colorlinks=true, allcolors=blue]{hyperref}
\usepackage{lipsum}
\usepackage{multirow}
\usepackage{xcolor}
\usepackage{orcidlink}
\usepackage{xspace}
\usepackage{soul}
\usepackage[switch]{linenoaa} 
\usepackage{txfonts}
\usepackage{natbib}
\usepackage{amsmath}    % Advanced maths commands
\usepackage{amssymb}    % Extra maths symbols
\newcommand{\onea}{A0620$-$00\xspace}
\DeclareMathOperator{\atantwo}{atan2}

\begin{document} 

\title{Optical and near-infrared polarization of the black hole X-ray binary \onea in quiescence} 

\titlerunning{Optical polarization of \onea}

\author{Vadim Kravtsov\inst{1}\orcidlink{0000-0002-7502-3173}
\and  
Alexandra Veledina\inst{1,2}\orcidlink{0000-0002-5767-7253}  
\and 
Andrei V. Berdyugin\inst{1}\orcidlink{0000-0002-9353-5164}  
\and
Juri Poutanen\inst{1}\orcidlink{0000-0002-0983-0049}
\and
Sergey~S.~ Tsygankov\inst{1}\orcidlink{0000-0002-9679-0793}
\and
Tariq~Shahbaz\inst{3,4}
\and
Manuel~A.~P.~Torres\inst{3,4}\orcidlink{0000-0002-5297-2683}
\and
Helen~E.~Jermak\inst{5}\orcidlink{0000-0002-1197-8501}
\and
Callum McCall\inst{5}\orcidlink{0000-0002-3375-3397}
\and
Iain~A.~Steele\inst{5}\orcidlink{0000-0001-8397-5759}
\and
Jari~J.~E.~Kajava\inst{1,6}\orcidlink{0000-0002-3010-8333}
\and
Vilppu~Piirola\inst{1}\orcidlink{0000-0003-0186-206X}
\and 
Takeshi Sakanoi\inst{7}   
\and 
Masato Kagitani\inst{7}   
\and 
Svetlana V. Berdyugina\inst{8,9}\orcidlink{0000-0002-2238-7416}
}

\authorrunning{V. Kravtsov et al.} 
 
\institute{Department of Physics and Astronomy, FI-20014 University of Turku, Finland\\
\email{vakrau@utu.fi}
\and
Nordita, KTH Royal Institute of Technology and Stockholm University, Hannes Alfv\'ens v\"ag 12, SE-10691 Stockholm, Sweden 
\and 
Instituto de Astrofísica de Canarias, E-38205 La Laguna, Tenerife, Spain
\and
Departamento de Astrofísica, Universidad de La Laguna, E-38206 La Laguna, Tenerife, Spain
\and
Astrophysics Research Institute, Liverpool John Moores University, 146 Brownlow Hill, Liverpool L3 5RF, United Kingdom
\and
Serco for the European Space Agency (ESA), European Space Astronomy Centre, Camino Bajo del Castillo s/n, E-28692 Villanueva de la Cañada, Madrid, Spain
\and
Graduate School of Sciences, Tohoku University, Aoba-ku,  980-8578 Sendai, Japan
\and
Istituto Ricerche Solari Aldo e Cele Daccò (IRSOL), Faculty of Informatics, Università della Svizzera italiana, 6605 Locarno, Switzerland
\and
Institut für Sonnenphysik (KIS), Georges-Köhler-Allee 401a, 79110 Freiburg, Germany
}
 
\abstract 
{We present simultaneous high-precision optical polarimetric and near-infrared (NIR) to ultraviolet (UV) photometric observations of the low-mass black hole X-ray binary \onea in a quiescent state. Subtracting interstellar polarization, estimated from a sample of field stars, we derived the intrinsic polarization of \onea. We show that the intrinsic polarization degree (PD) varies with the orbital period with an amplitude of $\sim0.3\%$, at least in the $R$ band, where the signal-to-noise ratio of our observations is the best. This implies that some fraction of the optical polarization is produced by a scattering of stellar radiation off the matter that follows the black hole in its orbital motion. In addition, we see a rotation of the orbit-average intrinsic polarization angle (PA) with the wavelength from $163\degr$ in the $R$ to $177\degr$ in the $B$ band. All of the above, combined with the historical NIR-to-optical polarimetric observations, demonstrates the complex behavior of the average intrinsic polarization of \onea: the PA continuously  rotates from the infrared to the blue band by $\sim53\degr$ in total, while the PD of~$\sim1\%$ remains nearly constant over the entire spectral range.  
The spectral dependence of the PA can be described by Faraday rotation with a rotation measure of $-0.2$\,rad\,$\mu$m$^{-2}$, implying a magnetic field of a few gauss in the  plasma surrounding the black hole accretion disk. 
However, our preferred interpretation of the peculiar wavelength dependence is an interplay between two polarized components with different PAs.
Polarimetric measurements in the UV range can help in distinguishing between these scenarios.
}

\maketitle 

%
%-------------------------------------------------------------------

\section{Introduction}

\onea is the prototypical black hole (BH) low-mass X-ray binary (LMXB). It was discovered during its 1975 outburst \citep{Elvis1975}, when its luminosity increased by more than a million times compared to the quiescent levels, reaching the Eddington luminosity.
Since then, the object has been in a quiescent state, with its spectrum dominated by the emission of a $0.4M_\odot$ K-type star orbiting a $\sim6M_\odot$ BH \citep{McClintock1986, vanGrunsven2017} with an orbital period ($P_{\rm orb}$) of $7.75$~h. 
Nevertheless, additional sources of nonstellar origin (such as jet, accretion disk, inner hot accretion flow, and hot spot) were considered in the literature to explain the excess of radio, infrared (IR), and ultraviolet (UV) emission of \onea near quiescence \citep{McClintock1995, Muno2006, Froning2011, Gallo2019, Cherepashchuk2019}. 
Their contributions to the broadband spectrum are difficult to distinguish, especially in light of the alteration between accretion modes (passive, loop, and active modes; \citealt{Cantrell2008}).
An additional source of information is needed to distinguish between different spectral models. 
Polarization of the optical radiation may provide such information.

Optical and near-infrared (NIR) radiation produced in various physical processes, including the electron scattering of stellar radiation off the accretion disk or flow and synchrotron emission in the presence of an ordered magnetic field, can be polarized. 
The polarization degree (PD), polarization angle (PA), and their spectral properties are different for different processes, which makes polarimetry a powerful technique for studies of the physical mechanisms responsible for the optical and NIR emission production in BH X-ray binaries. 
A recent systematic study has shown, however, that in many quiescent BHs, the intrinsic optical polarization (corrected for interstellar contribution) is very small -- in most cases $P_{\rm int} \lesssim 0.5\%$ \citep{Kravtsov2022a}.
On the other hand, several BH LMXBs show significant polarization during (or near) quiescence: MAXI~J1820+070 had a high PD  (up to $5\%$), with the blue spectrum and PA different from the jet direction, suggesting a BH spin-orbit misalignment \citep{Poutanen2022}; \onea had $\sim$1\% optical and NIR polarization in the quiescent state \citep{Dubus2008,Russell2016}. 

In X-ray quiescence, \onea shows optical state changes: according to the \cite{Cantrell2008} classification, there are two different states of optical activity –– passive and active. 
In the passive state, variations in the optical flux are consistent with ellipsoidal variations produced by the rotation of the tidally distorted optical companion. 
In the active state, the source is $20\%$ brighter and shows an aperiodic high-frequency variation usually called ``flickering.'' 
  
In this paper we present the results of quasi-simultaneous high-precision optical polarimetric and multiwavelength (NIR to UV) photometric observations of \onea during its passive quiescent state.
The paper is organized as follows. 
In Sect.~\ref{sect:data} we describe the observational data. 
In Sect.~\ref{sect:results} we present the main results of our study: determination of the intrinsic optical polarization of \onea, its significant orbital variability, and the rotation of the average intrinsic PA with wavelength.
In Sect.~\ref{sect:discussion} we discuss possible physical mechanisms that can reproduce the observed behavior. 
Finally, in Sect.~\ref{sect:conclusions} we summarize our findings.

\section{Data acquisition and analysis}
\label{sect:data}

\subsection{Optical polarimetric observations}

High-precision optical polarimetric observations of \onea were performed using broadband \textit{BVR} polarimeter DIPol-UF  \citep{Piirola2020a}, a visitor instrument installed at the 2.56 m Nordic Optical Telescope (NOT), Observatorio del Roque de los Muchachos, La Palma, Spain.
Field stars used for the interstellar (IS) polarization estimation were observed with DIPol-2 \citep{Piirola2014}, mounted on the remotely controlled 60 cm Tohoku telescope (T60) at Haleakala Observatory, Hawaii.
Both polarimeters utilize a ``double-image'' design that effectively eliminates the polarization of the sky, even if it varies throughout the observations.
The instrumental polarization of both instruments is small ($<10^{-4}$) and was well calibrated by observing of 10--15 unpolarized standard stars. 
The zero points of the PAs were determined by observing highly polarized standards HD~236928 and HD~25443. 
A more detailed description of the methods and calibrations can be found in \citet{Piirola2020} and  \citet{Kravtsov2022a}, and references therein. 

\onea was observed on two nights between 2022 December 24 and 26, and 140 and 52 individual measurements of the Stokes parameters were made on the two nights, respectively.
The first observation was continuous, 8~h long, and hence covered the whole orbital period of the binary, while the second observation covered only 40\% of the period.
Hereafter we refer to the average polarization measured during the first observation as the orbit-averaged polarization. 
Each individual linear polarization measurement was obtained from four consecutive images with 50-s exposures taken at different half-wave plate positions, resulting in one polarization measurement per $\sim$3.3~min.
To increase the signal-to-noise ratio, we split the data into bins such that each bin contains ten individual measurements of the Stokes parameters.
The errors of the normalized Stokes parameters $q$ and $u$ were computed as the standard errors of the weighted mean values within the bin. 
The Stokes parameters $(q, u)$ then were translated into the PD ($P$) and PA ($\theta$),
\begin{equation}
P = \sqrt{q^2 + u^2},\qquad 
\theta = \frac{1}{2}\atantwo(u, q).  
\label{eq:pd_pa}
\end{equation}
The uncertainty on the PD is equal to the uncertainty of the individual Stokes parameters, and uncertainty on the PA in radians was estimated as $\sigma_{\theta} = \sigma_{P}/(2P)$ \citep{Serkowski1962,Kosenkov2017}. The phase-resolved PD was corrected for the bias caused by the low signal-to-noise ratio using the relation $P_0 = (P^2 - 2\sigma^2_P)^{1/2}$ \citep{Simmons1985}.

\begin{table}%[]
\centering
\caption{Polarimetric and photometric observations of \onea in December 2022.}
\begin{tabular}{llll}
\hline\hline
Telescope & UT Date 2022 & Filters & $N_\textrm{obs}$ \\
\hline
NOT  &  Dec 24--26   &   \textit{B, V, R}                    & 2  \\
GTC  &  Dec 25       &   $K_\textrm{s}$, \textit{H}, \textit{J}                 & 1  \\
LT   &  Dec 20--24   &   \textit{u, g, r, i, z,} \textit{B, V} & 3  \\
UVOT &  Dec 23--26   &   \textit{B, V, u}, \textit{w}1, \textit{m}2, \textit{w}2         & 4  \\
\hline
\end{tabular}
\label{tab:obs_log}
\end{table}

\subsection{Multiwavelength photometry}

Quasi-simultaneous multiwavelength photometric observational campaign was organized on several telescopes (see Table~\ref{tab:obs_log}). 
Near-infrared \textit{JHK} photometry was made using EMIR wide-field imager \citep{EMIR}, installed on the 10.4\,m Gran Telescopio Canarias (GTC), La Palma, Spain.  
Observations in the broadband SDSS-\textit{ugriz}, Bessel \textit{V}, and Bessel \textit{B} filters were performed using IO:O instrument of the 2-m Liverpool Telescope \citep[LT;][]{Steele04}, La Palma, Spain. 
For all the instruments, basic data reductions such as bias and dark subtraction and flat fielding are done via the internal common pipelines.
As the photometric standards, we used two stars with known Sloan Digital Sky Survey (SDSS) magnitudes. To obtain fluxes of the object from its magnitudes we used standard zero-points for SDSS and Johnson-Cousins systems \citep{Fukugita96, Bessel79}.

\subsection{Swift/UVOT}

\textit{The Neil Gehrels Swift} Observatory \citep{2004ApJ...611.1005G} observed \onea four times from 2022 December 23 to 26 (see Table~\ref{tab:obs_log}) with total exposure of about 7\,ks. The image analysis has been done following the procedure provided by the UK Swift Science Data Centre.\footnote{\url{https://www.swift.ac.uk/analysis/uvot/}} 
Photometry in all available filters ($V$, $B$, $U$, $UVW1$, $UVW2$, and $UVM2$) was performed using the tool {\sc uvotsource} from the {\sc heasoft} package version 6.32 and the latest calibration files. The source and background photons were extracted from the apertures with radii of 5\arcsec\ and 10\arcsec, respectively. The background was chosen with the center about 18\arcsec\ away from the source for all filters.

For the spectral fitting, all available data were converted to the spectral files. For the \textit{Swift}/UVOT data the {\sc uvot2pha} tool was applied using the corresponding response files in the CALDB. All other data were converted to the spectral files using tool {\sc ftflx2xsp} from the {\sc ftools} package. The following spectral fitting was performed using {\sc xspec} version 12.13.1 \citep{Arnaud96}.

\section{Results}
\label{sect:results}

\subsection{Average intrinsic polarization}

\begin{figure}
\centering
\includegraphics[width=0.89\linewidth]{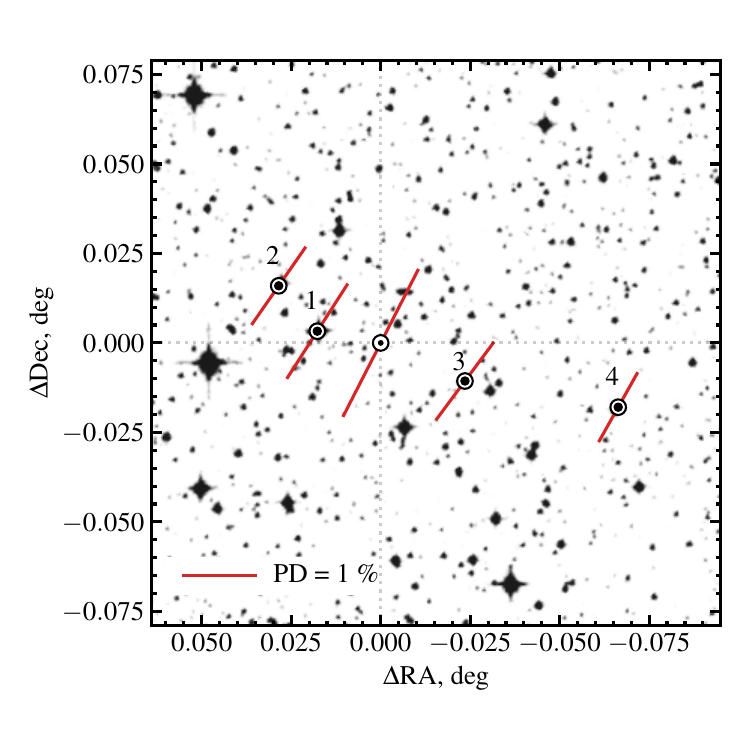}
\caption{Polarization map of \onea (at the origin) and field stars in the $R$ band. The lines correspond to the observed polarization, with the length of the bars showing the PD, and the direction indicating the PA (measured from north to east).}
\label{fig:ispolmap}
\end{figure}
 
\begin{figure}
\includegraphics[width=0.8\linewidth]{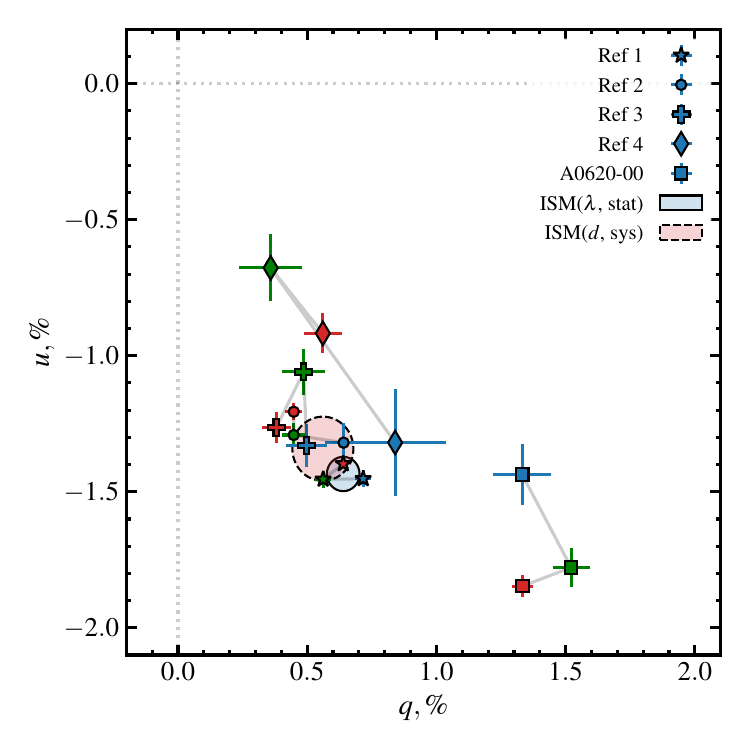}
\caption{Observed orbit-average Stokes parameters of \onea (squares) and field stars in the $B$, $V$, and $R$ bands (blue, green, and red markers, respectively). The light blue circle shows the uncertainty of the IS polarization with wavelength $\sigma_{\rm is, \lambda}$; the light red circle shows the systematic error on the IS polarization zero point, $\sigma_{\rm is, sys}$.}
\label{fig:quall}
\end{figure}

\begin{table*}
\centering
\caption{Polarization of field stars and \onea.}
\label{table:ref_pol}
\begin{tabular}{lc*2{r@{\,$\pm$\,}l}c*2{r@{\,$\pm$\,}l}c*2{r@{\,$\pm$\,}l}}
        \hline
        \hline
        & &  \multicolumn{4}{c}{$B$} & & \multicolumn{4}{c}{$V$} && \multicolumn{4}{c}{$R$} \\
\cline{3-6} 
\cline{8-11}
\cline{13-16} 
        Field  star & Parallax (mas) & \multicolumn{2}{c}{$q$ (\%)} & \multicolumn{2}{c}{$u$ (\%)} && \multicolumn{2}{c}{$q$ (\%)} & \multicolumn{2}{c}{$u$ (\%)} && \multicolumn{2}{c}{$q$ (\%)} & \multicolumn{2}{c}{$u$ (\%)} \\
        \hline
%        \hline
        Ref 1  & $0.72\pm0.04$  & 0.71 & 0.03 & $-1.45$ & 0.03 && 0.56 & 0.03 & $-1.45$ & 0.03 && 0.64 & 0.02 & $-1.40$ & 0.02 \\
        Ref 2  & $0.87\pm0.02$  & 0.64 & 0.07 & $-1.32$ & 0.07 && 0.45 & 0.04 & $-1.29$ & 0.04 && 0.45 & 0.03 & $-1.20$ & 0.03 \\
        Ref 3  & $0.56\pm0.02$  & 0.50 & 0.08 & $-1.33$ & 0.08 && 0.49 & 0.08 & $-1.06$ & 0.08 && 0.38 & 0.06 & $-1.26$ & 0.06 \\
        Ref 4  & $0.55\pm0.02$  & 0.84 & 0.20 & $-1.32$ & 0.20 && 0.36 & 0.12 & $-0.68$ & 0.12 && 0.56 & 0.07 & $-0.92$ & 0.07 \\
        \onea  & $0.69\pm0.12$  & 1.33 & 0.11 & $-1.44$ & 0.11 && 1.52 & 0.07 & $-1.78$ & 0.07 && 1.33 & 0.04 & $-1.85$ & 0.04 \\
        
        \hline
   
    \end{tabular}
    
\end{table*}

\begin{figure}
\includegraphics[width=0.8\linewidth]{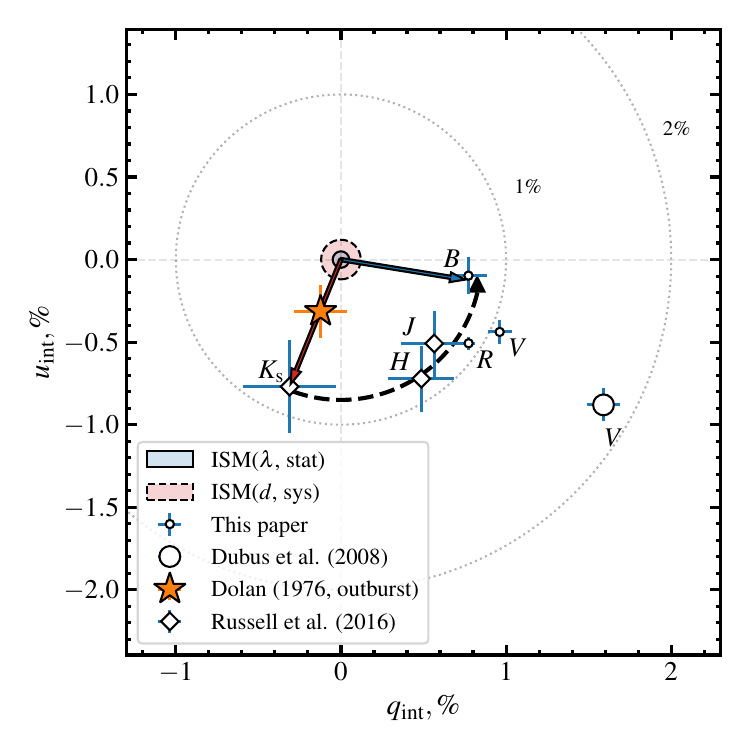}
\caption{Intrinsic Stokes parameters of the polarization of \onea in quiescence measured in different filters from the $K_{\rm s}$ to $B$, as indicated by letters near each data point. The light blue circle at the origin illustrates the possible uncertainty of the IS polarization with wavelength, $\sigma_{\rm is, \lambda}$; the light red circle shows the systematic error on the IS polarization zero point, $\sigma_{\rm is, sys}$. The red and blue arrows indicate the intrinsic polarization vector directions in the $K_{\rm s}$ and $B$ bands, respectively. The dashed curved arrow shows the track left by the intrinsic polarization vector during its rotation from the IR to the blue part of the spectrum. An orange star shows the polarization of \onea in $V$ measured during the 1975 outburst by \citet{Dolan1976}.}
\label{fig:intrinsic_qu}
\end{figure}

The observed optical polarization of a distant star is a combination of the intrinsic polarization of the object and IS polarization.
The main source of the IS polarization  ($q_{\rm is}$, $u_{\rm is}$) is the optical dichroism of nonspherical IS dust particles aligned by Galactic magnetic field. 
The observed polarization of \onea is dominated by the IS component, as evidenced by the rough alignment (within 10\degr; see Fig.~\ref{fig:ispolmap}) of the observed PA of the source with those of the field stars.
To extract the intrinsic polarization ($q_{\rm int}$, $u_{\rm int}$) of \onea from the observed polarization ($q_{\rm obs}$, $u_{\rm obs}$), we first estimated the IS polarization. 
IS polarization usually depends on the distance to the source and may be different at different wavelengths. 
To study it, we observed four close field stars with similar \textit{Gaia} DR3 \citep{GaiaDR3} parallaxes  in the \textit{BVR} bands (see Table~\ref{table:ref_pol}). 
The observations of the brightest field star (Ref~1) show very little wavelength dependence of IS polarization in the direction to \onea~ in the optical band -- it is consistent with being constant with wavelength at the level of better than $\sigma_{\rm is, \lambda} = 0.05\%$ (blue circle in Figs.~\ref{fig:quall}  and \ref{fig:intrinsic_qu}). 
Thus, one zero point ($q_{\rm is}$, $u_{\rm is}$) for IS polarization can be used for all three optical bands. 
Due to the possible dependence on distance, the exact position of this zero point on the $qu$-plane is uncertain.
Different approaches can be used to set this zero point: one can use either the Stokes parameters of the closest in distance field star or the weighted average Stokes parameters of the group of field stars located at similar distances. 
We used the later approach to find ($q_{\rm is}$, $u_{\rm is}$) based on the observations of the field stars Ref 1 -- Ref 3, for which the accuracy of our observations was the best. 
The scatter in their Stokes parameters determines the systematic error on the IS polarization zero point $\sigma^{\rm sys}_{\rm is} = 0.12\%$ (shown as red circle in Figs.~\ref{fig:quall}  and \ref{fig:intrinsic_qu}).
To find the intrinsic polarization of \onea, we subtracted the IS Stokes parameters ($q_{\rm is}$, $u_{\rm is}$) from the observed Stokes parameters ($q_{\rm obs}$, $u_{\rm obs}$) of the source. 
The statistical errors on the intrinsic Stokes parameters of \onea were calculated as $\sigma^2_{\rm int} = \sigma^2_{\rm obs} + \sigma^2_{\rm is, \lambda}$; the $\sigma^2_{\rm is, \lambda}$ term includes possible dependence of the IS polarization on wavelength. 
Then, Eq.~(\ref{eq:pd_pa}) is used to calculate the PD and PA together with their statistical errors.
The systematic error $\sigma_{\rm is, sys}$ on the IS zero point translates to the systematic errors of the intrinsic PD and PA $\sigma^{\rm sys}_{p} = 0.12\%$ and $\sigma^{\rm sys}_{\theta} \approx 4\degr$.

\begin{table*}   
\caption{One-orbit-average observed and intrinsic PD and PA of \onea together with the IS polarization estimate.  
}             
\label{table:polar}      
\centering                        
\begin{tabular}{ccccccccc}          
\hline\hline                
  & \multicolumn{2}{c}{Observed} & & \multicolumn{2}{c}{Interstellar} & &  \multicolumn{2}{c}{Intrinsic}  \\  
\cline{2-3} 
\cline{5-6}
\cline{8-9}
F & $P_{\text{obs}}$ (\%) & $\theta_{\text{obs}}$ (deg) & & $P_{\text{is}}$ (\%)  & $\theta_{\text{is}}$ (deg) && $P_{\text{int}}$ (\%) & $\theta_{\text{int}}$ (deg)  \\   
\hline                        
   \textit{B}  & $1.96 \pm 0.11$ & $156 \pm 2$ &&  &  && $0.78 \pm 0.12^{\rm stat} \pm 0.12^{\rm sys}$ & $177 \pm 5^{\rm stat} \pm 5^{\rm sys}$ \\ 
   \textit{V}  & $2.34 \pm 0.07$ & $155 \pm 1$ && $1.45 \pm 0.05^{\rm stat} \pm 0.12^{\rm sys}$ & $146 \pm 1^{\rm stat} \pm2^{\rm sys}$ && $1.05 \pm 0.09^{\rm stat}\pm 0.12^{\rm sys}$ & $168 \pm 2^{\rm stat} \pm 3^{\rm sys}$ \\
   \textit{R}  & $2.27 \pm 0.04$ & $153 \pm 1$ &&  &  && $0.92 \pm 0.06^{\rm stat}\pm 0.12^{\rm sys}$ & $163 \pm 2^{\rm stat} \pm 4^{\rm sys} $ \\ 
\hline                                   
\end{tabular}
\end{table*}

\begin{figure}
\centering
\includegraphics[width=0.85\linewidth]{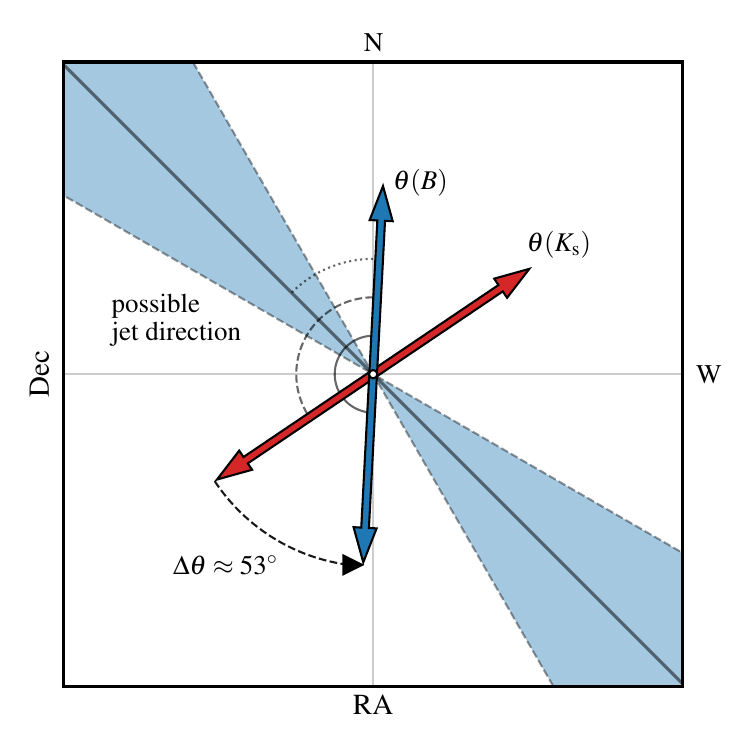}
\caption{Rotation of the polarization on the sky. Red and blue arrows correspond to the direction of the polarization in the $B$ and $K_{\rm s}$ filters, respectively. The blue region shows the direction of the radio ejections as measured in \citet{Kuulkers1999}.}
\label{fig:rotation_on_the_sky}
\end{figure}

Subtracting the IS component from the observed polarization, we find the average intrinsic polarization of \onea to be $P = $~0.8--1.0\% with $\theta$ changing from $163\degr$ in the $R$ filter to $177\degr$ in $B$ (see Table~\ref{table:polar}). The fit with a constant to the intrinsic PA gives unacceptable fit with $\chi^2/\rm{d.o.f.}$=10/2.
Combining NIR polarimetric observations \citep{Russell2016} corrected for IS polarization with our optical measurements, we see the continuous rotation of the intrinsic polarization vector with the wavelength from NIR to $B$ (see Fig.~\ref{fig:intrinsic_qu}).
The fit with a constant to all intrinsic PAs from NIR to B gives unacceptable fit with $\chi^2/\rm{d.o.f.}$=32/6.
The amplitude of the polarization vector rotation on the sky is $\Delta \theta = \theta(B) - \theta(K_{\rm s}) \approx  53\degr$ (see Fig.~\ref{fig:rotation_on_the_sky}). In contrast to the PA, PD barely changes with wavelength remaining at $\sim$1\% level from $K_{\rm s}$ to $B$ band (constant fit gives $\chi^2/\rm{d.o.f.}$=5/6). 

We note that although the determination of the intrinsic PA strongly depends on the accuracy of the IS polarization estimate, there is an additional reason to believe that the intrinsic polarization estimate is close to its true value.
Although the observed PD of \onea in \citet{Dubus2008} $V$-band observations was significantly higher than in this paper, subtracting our estimate of the IS polarization from their observed Stokes parameters, we get the PA of intrinsic polarization matching our value within a few degrees -- this can be seen in Fig.~\ref{fig:intrinsic_qu}, if one connects the origin to the corresponding observational points and compare the directions of the resulting vectors. 
This alignment of the intrinsic polarization vectors is unlikely to be coincidental and may appear naturally if the intrinsic PD of the source changes while the PA is stable.

\begin{figure}
    \centering    \includegraphics[width=0.9\linewidth]{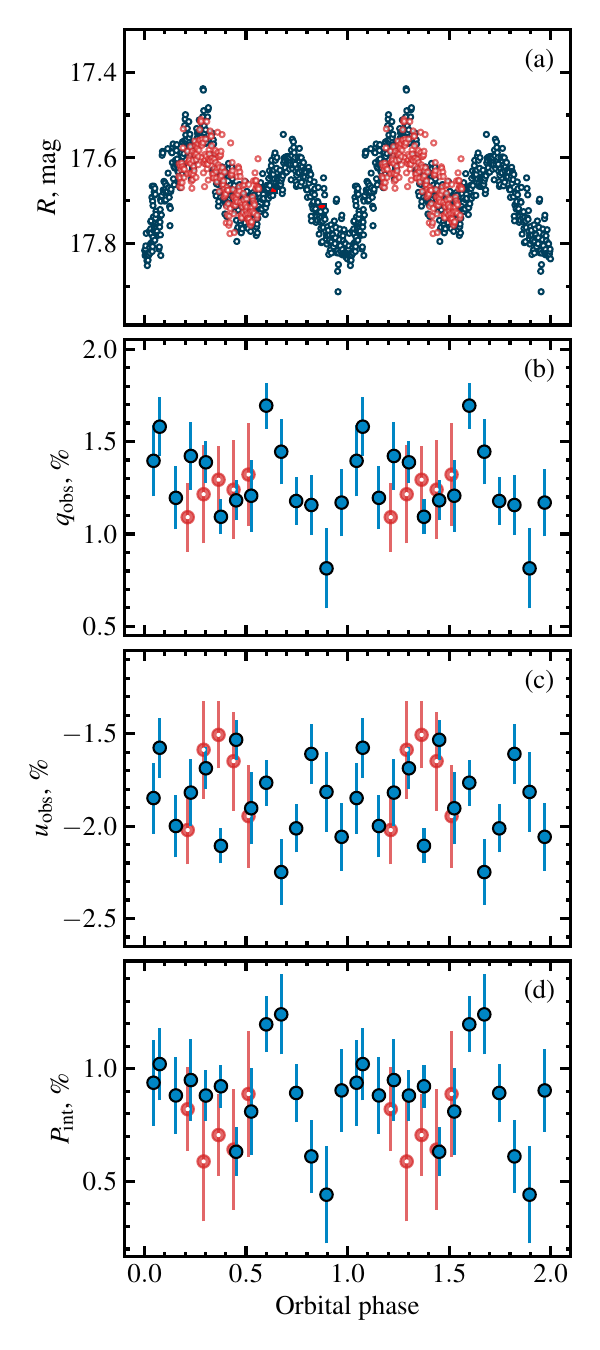}
\caption{Orbital profiles of the flux and of the normalized Stokes parameters of \onea. Solid blue and empty red circles correspond to the data from two different orbits. \textit{Panel (a):} Photometric $R$ magnitudes of \onea  folded with the orbital period. 
\textit{Panels (b)} and \textit{(c):}  Observed normalized Stokes parameters. 
\textit{Panel (d):}  Intrinsic PD of \onea in the $R$ filter folded with the orbital period. Each circle with a 1$\sigma$ error bar shows the 35-minute average polarization. }
\label{fig:phot_and_pol} 
\end{figure}

\subsection{Orbital variability of the polarization}
We folded the photometric and polarimetric observations of \onea in the $R$ band (for which the S/N is the best) with the orbital period using the recent ephemeris of \citet{Cherepashchuk2019}. 
We find significant variability of intrinsic polarization of the source in the $R$ filter (Fig.~\ref{fig:phot_and_pol}b,c).
We plotted the orbital profile of the observed polarization together with the optical light curve of \onea, obtained simultaneously (Fig.~\ref{fig:phot_and_pol}a). 
We see the pronounced peak in PD at the orbital phase of 0.75 and two minima at phases around 0.5 and 1.0.
Similar peak in PD around phase 0.7 has been observed by \citet{Dolan1989}.
Such polarization behavior with minima in conjunctions and maxima in quadratures is expected and has been observed in other binary systems (discussed in detail in Sect.~\ref{sect:variability}), and although our observations covered only 1.5 orbits, and the S/N for individual bins is at the threshold of significance, we can still cautiously state that the nature of observed polarization variability is more likely to be orbital rather than stochastic.
The absence of flickering in the photometric observations together with the visual magnitude $V = 18.3 \pm 0.1$ of \onea suggests that the source was in the passive quiescent state during our campaign \citep{Cantrell2008}.

\subsection{Broadband spectrum}

\begin{figure}
\includegraphics[width=0.97\linewidth]{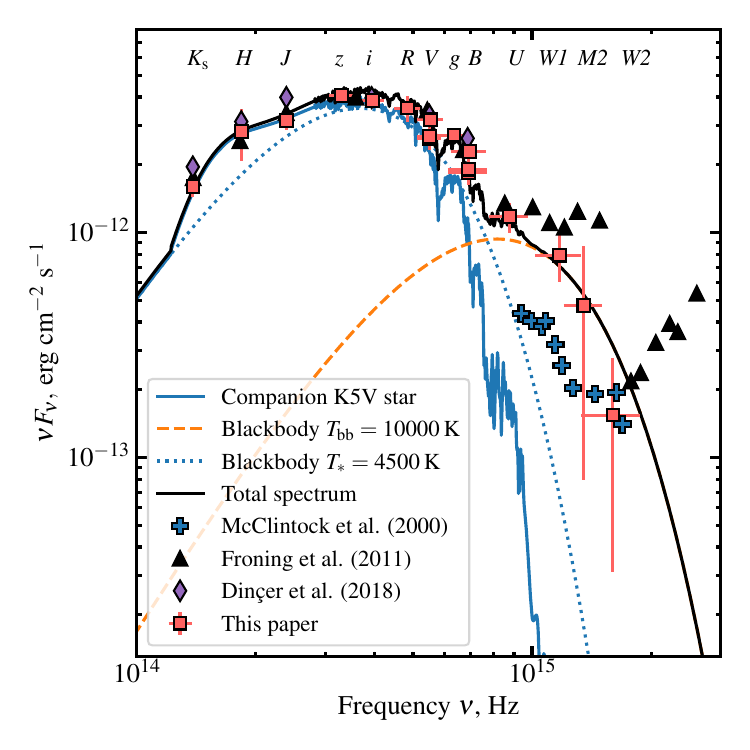}
\caption{SED of \onea, corrected for reddening assuming $E(B - V) = 0.35$. Red squares with error bars correspond to the data from this paper. Blue crosses, black triangles, and purple diamonds show data from \citet{McClintock2000}, \citet{Froning2011}, and  \citet{Dincer2018}, respectively. Solid blue, dotted blue, and dashed orange lines show the spectra of the K5V star \citep{Pickles1998}, the blackbody of $T_{*} = 4500$\,K, and an additional blackbody of $T_{\rm bb} = 10,000$\,K, respectively.}
    \label{fig:spectrum}
\end{figure}

The observed NIR-to-UV spectral energy distribution (SED) of \onea in passive quiescent state, corrected for reddening assuming $E(B - V) = 0.35$ \citep{Wu1983} is shown in Fig.~\ref{fig:spectrum}.
The key feature of the SED is the excess of UV photons, previously observed in the quiescent state \citep{Froning2011, McClintock1995} -- an extra source of UV radiation in addition to the optical companion is needed to reproduce the observed shape of the spectrum.
The spectrum alone, however, does not allow an unambiguous separation of the second component: even the simplest model with one additional blackbody is degenerate due to the uncertainty in the stellar spectrum normalization.
Various different methods were used to estimate the contribution of nonstellar emission to the quiescent spectrum of \onea \citep{Marsh1994, Gelino2001, Harrison2007, Froning2011, Dincer2018}. 
Despite the fact that there is no complete consensus, it can be cautiously noted that most authors agree that the companion star dominates the entire NIR-to-optical range in the passive quiescent state, with the contribution of additional component increasing toward the blue part of the spectrum. 
The additional component becomes comparable with the stellar flux only near the $B$ band, where its contribution reaches 20--50\%. 
Nonstellar emission spectrum can be described with the blackbody of temperature $T_{\rm bb}=$ 9000--11000\,K  depending on the state of activity.
The nature of the additional component is still under discussion, but most likely it corresponds to the brightest part of the accretion disk (either the hot inner disk regions or the bright spot formed at the impact point of the accretion stream). 
The bright spot is clearly present in the Doppler tomograms \citep{Marsh1994, Shahbaz1994, Shahbaz2004, Neilsen2008} and it is needed to explain the asymmetric light curves \citep{Froning2001, Cantrell2010, vanGrunsven2017, Cherepashchuk2019}. 
The contribution of the nonstellar radiation is variable on short \citep{Haswell1993} and long \citep{Cantrell2008} timescales. 
Our new observations are consistent with both interpretations (see Fig.~\ref{fig:spectrum} and Sect.~\ref{sect:discussion}).

\section{Discussion}
\label{sect:discussion}

\subsection{The source of the polarization}
\label{sect:source_of_polarization}

The first optical polarimetric observations of \onea were performed during its 1975 outburst \citep[][see the orange star in Fig.~\ref{fig:intrinsic_qu}]{Dolan1976}.
The observed PD, PA, and their spectral properties were found to be consistent with the IS origin.
Our IS polarization estimate is very close to the outburst polarization level of \onea, which confirms the above statement.
Sub-percent intrinsic optical polarization of \onea is in line with that of the other BH X-ray binaries -- most known sources observed during the outburst show low intrinsic polarization levels (\mbox{XTE~J1118+480}, \citealt{Schultz2004};   \mbox{MAXI~J0637$-$430}, \citealt{Kravtsov2019ATel}; \mbox{MAXI~J1820+070}, \citealt{Veledina2019, Kosenkov2020b};  \mbox{Swift~J1727.8$-$1613}, \citealt{Kravtsov2023ATel}; \mbox{V404~Cyg}, \citealt{Tanaka2016,Kosenkov2017};  \mbox{GX~339$-$4}, \citealt{Mastroserio2025}; \mbox{LMC~X-3}, \citealt{Boyd2001}). 
However, the intrinsic PD of some BH LMXBs  increases significantly  as they approach quiescence (e.g., \mbox{MAXI~J1820+070}, \citealt{Poutanen2022}, and \onea, \citealt{Dubus2008}). 
Intrinsic optical polarization is also detected in some NS X-ray binaries (\mbox{Aql~X-1}, \citealt{Charles1980}; \mbox{Cyg~X-2}, \citealt{KochMiramond1995}; \mbox{Sco~X-1}, \citealt{Schultz2004}; \mbox{4U~0614+091}, \citealt{Baglio2014}; \mbox{XTE~J1709--267}, \citealt{Higgins2019}).

The orbit-averaged intrinsic polarization of \onea in the quiescent state has nearly the same PD~$\sim 1\%$ at all wavelengths from the $K_{\rm s}$ to $B$ band, yet PA shows a significant rotation over those wavelengths (Fig.~\ref{fig:intrinsic_qu}). 
This makes \onea the only source known to date for which such a strong PA dependence on the wavelength is observed. 
Such polarization behavior cannot be explained in terms of simple mechanisms of polarization production. 
Indeed, if the polarization is produced mostly by scattering, its PA should not depend on the wavelength.

\begin{figure}
\centering
\includegraphics[width=\linewidth]{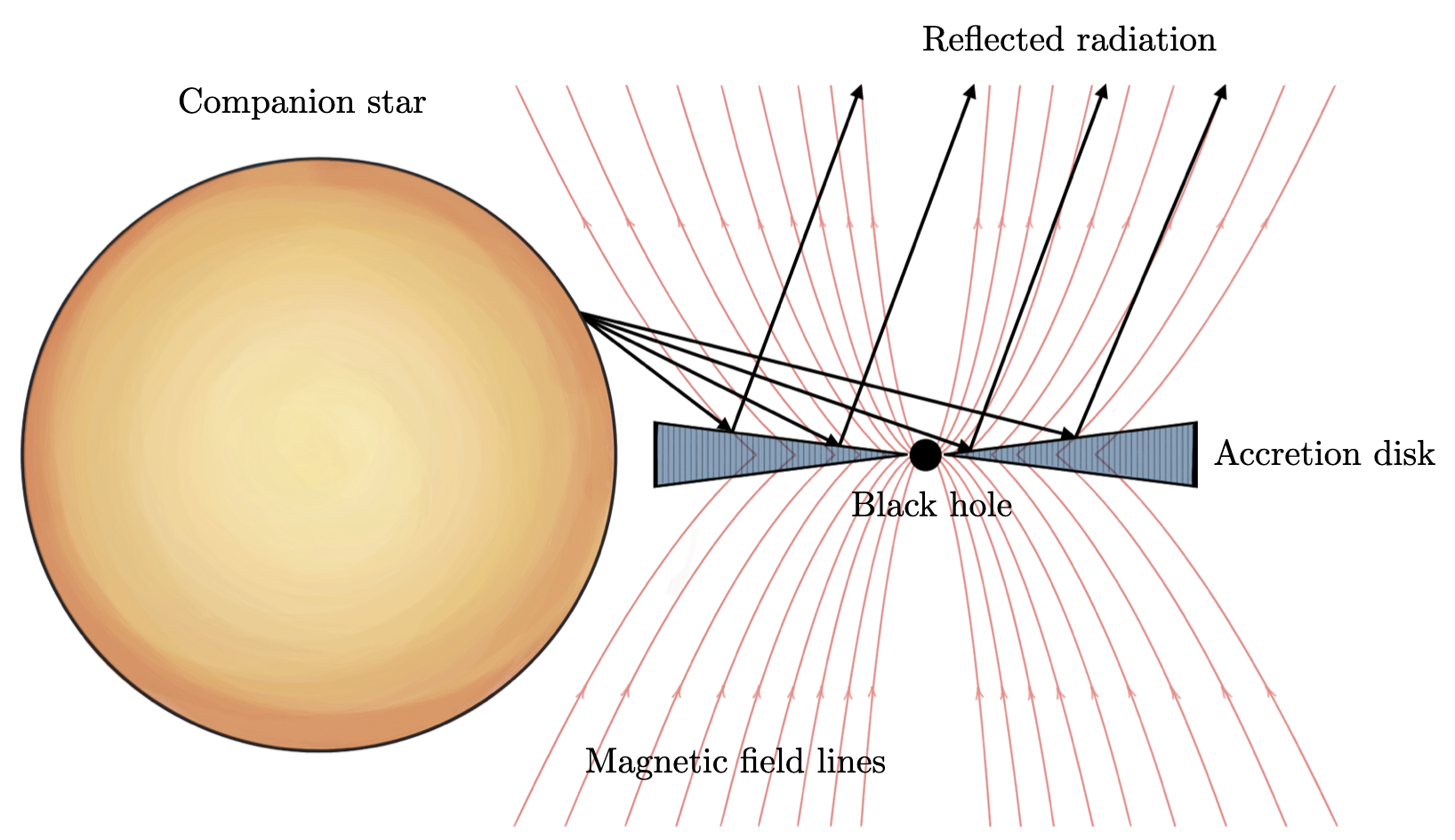}
\caption{Schematic illustration of the scattering geometry (not to scale).}
\label{fig:sketch}
\end{figure}

\begin{figure}
\includegraphics[width=0.9\linewidth]{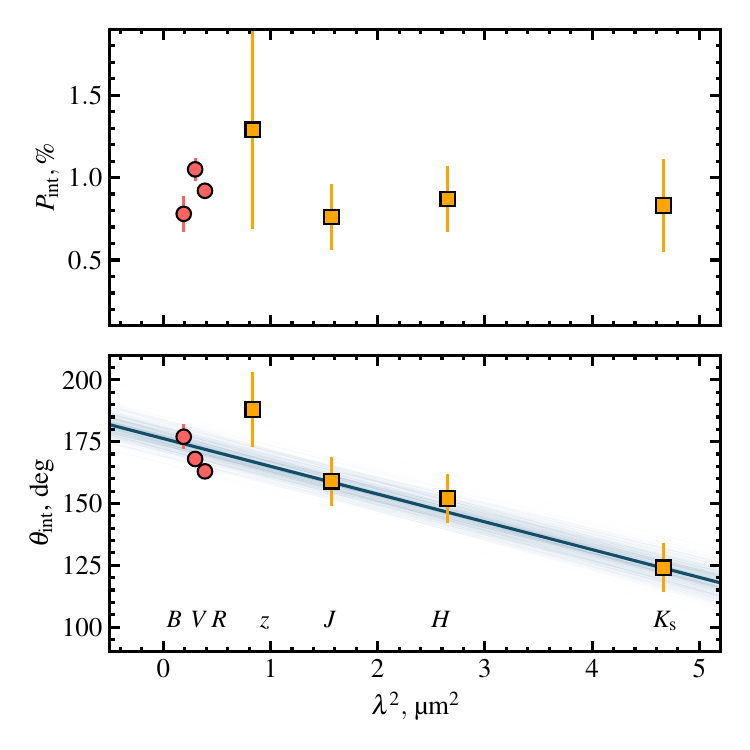}
\caption{Dependence of the intrinsic PD and PA of \onea on $\lambda^2$. The circles and squares correspond to optical observations from this paper and NIR polarimetric observations from \citet{Russell2016}, respectively. The blue line shows the best-fit Faraday rotation model given by Eq.\,\eqref{eq:Faraday}.}
\label{fig:Faraday}
\end{figure}

We first studied the case in which the observed dependence of PA on wavelength is caused by Faraday rotation. 
We considered the scenario illustrated in Fig.~\ref{fig:sketch}. 
The unpolarized radiation of the companion star gets scattered by the accretion disk and the polarized scattered radiation has initial PA independent of the wavelength.
After the scattering, the linearly polarized radiation propagates toward the observer through the magnetized plasma surrounding the accretion disk.
The PA of the light propagating along the magnetic field lines experiences the Faraday rotation, resulting in the observed PA dependence on wavelength.

The PA of linearly polarized radiation propagating through the magnetized plasma rotates with the wavelength $\lambda$ as
\begin{equation}
\theta(\lambda) = \theta_0 + {\rm RM}\,\lambda^2,     
\label{eq:Faraday}
\end{equation}
where the rotation measure (RM) is defined through the integral along the line of sight: 
\begin{equation}
\mbox{RM} = \frac{e^3}{2\pi m_{\rm e}^2 c^4} \int {n_{\rm e}(l) B_{||}(l)dl} , 
\end{equation} 
where $n_{\rm e}$ is the electron number density, $B_{||}$ (in~G) is the line of sight magnetic field strength, $e$ is the electric charge, $m_{\rm e}$ is the electron mass and $c$ is the speed of light. 
If $B_{||}$ is constant, the RM can be expressed through the Thomson optical depth $\tau_{\rm T}=\int n_{\rm e} \sigma_{\rm T} dl$. 
In this case, RM$\approx 0.4\tau_{\rm T} B_{||}$\,rad\,$\mu$m$^{-2}$.

We used Eq.~\eqref{eq:Faraday} with parameters $\theta_0$ and RM to find the best-fit solution for the Faraday rotation of PA in our optical points and NIR points reported in \citet{Russell2016}.
We find the best-fit parameters $\theta_0=175\degr \pm 4\degr$ and RM=$-0.20\pm0.01$\,rad\,$\mu$m$^{-2}$ (see Fig.~\ref{fig:Faraday}).
The latter value gives the relation between the line-of-sight optical depth and magnetic field (in gauss) $\tau_{\rm T}B_{||} \approx0.5$.
For the realistic values of matter density in quiescence $\tau_{\rm T}\lesssim0.1$, we find $B\gtrsim5$~G.
This value is roughly consistent with the magnetic fields expected in the quiescent-state optically thin accretion flows \citep{Wallace2021}, which may be similar to those of optically thin plasma surrounding the disk (Fig.~\ref{fig:sketch}).

We note, however, that the NIR data were taken almost ten years prior to our optical observations, and hence potential magnetic field variations may lead to the inconsistency of trends between the optical and NIR PA points.
We indeed observe that our recent measurements (leftmost three points in Fig.~\ref{fig:Faraday}) lie on a straight line with the slope differing from the general trend.
If we apply Eq~\eqref{eq:Faraday} only to our data, we obtain $\theta_0=186\degr \pm 6\degr$ and RM=$-1.3\pm0.3$\,rad\,$\mu$m$^{-2}$ (see Fig.~\ref{fig:Faraday}).
This translates to $\tau_{\rm T} B_{||}\approx3.2$, requiring almost seven times higher magnetic field strength for the Faraday rotation to occur in the optically thin plasma ($\tau_{\rm T}\lesssim1$).

\begin{figure}
\centering  \includegraphics[width=0.9\linewidth]{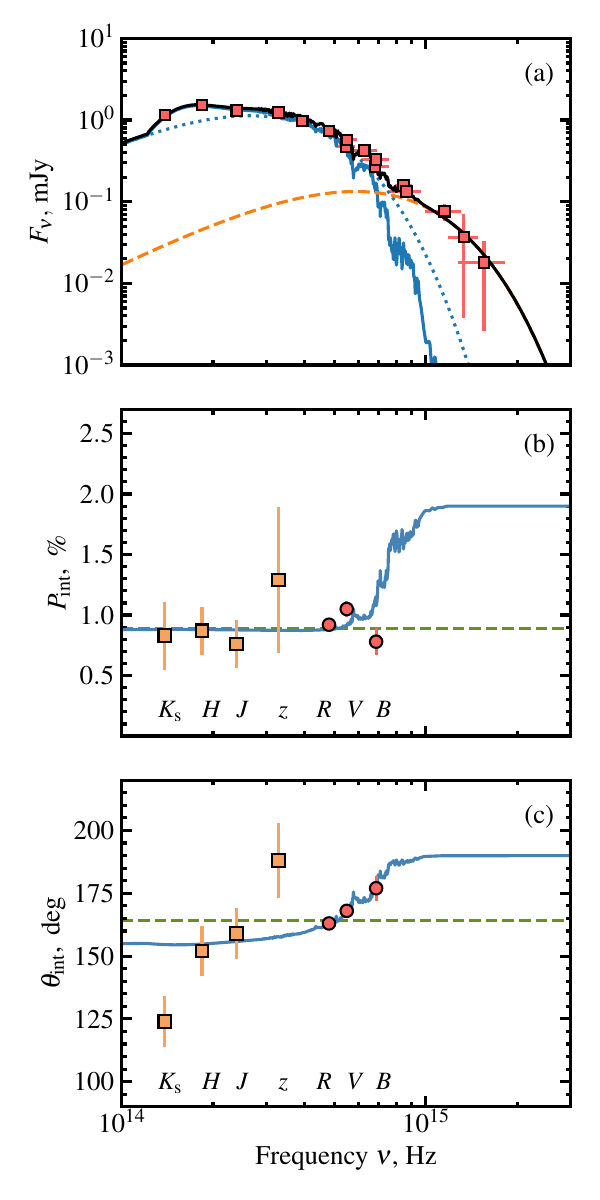}
\caption{Polarimetric properties of \onea. 
\textit{Panel (a):} Energy dependence of flux ($F_\nu$), as in Fig.~\ref{fig:spectrum} but only the data from this paper are shown. \textit{Panels (b)} and \textit{(c):} Energy dependence of the intrinsic PD and PA of \onea. Circles and squares correspond to optical observations from this paper and NIR polarimetric observations from \citet{Russell2016}, respectively. The blue line corresponds to the model with two polarized components described in Sect.~\ref{sect:source_of_polarization}. Dashed green lines correspond to the constant polarization model.}
    \label{fig:sed_model}
\end{figure}

Next, we considered the scenario in which the observed polarization arises from two components: one coming from the scattering of the stellar light and the other from the polarized contribution of the additional UV component.
The scattering component is assumed to have constant PD and PA, which implies a constant fraction of the scattered radiation, and spectrum as that of the optical companion (solid blue line in Fig.~\ref{fig:sed_model}a).
The additional UV component (dashed orange line), which was used to describe the UV part of the observed SED, is set to have constant PD and PA, whose values were free to vary. 
The blue lines in Fig.~\ref{fig:sed_model}b and c show the dependence of PD and PA for this two-component scenario.
To reproduce the rotation of the PA from the NIR to the $B$ band, the components must have different PAs: PA$_{\rm sc}\approx$150\degr\ and PA$_{\rm UV}\approx$190\degr.
Because the components have comparable contribution to the optical flux, the PD of the UV component (PD$_{\rm UV}\approx$1.8\%) should be significantly higher than that of the scattering component (PD$_{\rm sc}\approx$0.8\%) to dominate the PA in~$B$. The polarization of the UV component can arise from the first (single) Compton up-scattering of the disk or synchrotron photons by electrons in the hot accretion flow, similar to the case of MAXI~J1820+070 \citep{Poutanen2022}.
The difference of PA in the scattered and UV components can then indicate misalignment by $\sim$50\degr\ (or $\sim$40\degr) between the axes of these components.
This scenario predicts rise of PD and further rotation of PA toward the UV band (see Fig.~\ref{fig:sed_model}).

While this model fairly well describes the rotation of the optical PA, it cannot reproduce the dataset in full detail. 
First of all, the intrinsic PAs in $K_{\rm s}$ and $z$ differ from the model predictions. 
This difference may arise due to the fact that the observations in NIR were preformed a decade prior to the optical campaign while the source may have been in a different state (loop or active, \citealt{Cantrell2008}). 
There is evidence that in active quiescent state, the additional component contributes to the NIR part of the spectrum \citep{Froning2007, Dincer2018, Cherepashchuk2019}, which could be a jet, an accretion stream-disk impact point, or a dusty circumbinary disk \citep{Muno2006, Gallo2007, Cherepashchuk2019}. 
This additional NIR component, not present in our passive state observations, could however be responsible for an additional polarization in the $K_{\rm s}$ band in the epoch of the \cite{Russell2016} observations; that would explain the observed difference in the PA.
Some deviations in NIR may also arise from the Faraday rotation.
Additionally, although the model describes the flat PD spectrum from NIR to $R$ reasonably well  and replicates the PD rise due to the absorption line in $V$ well, the PD in $B$ is about 0.3\% lower than the level predicted by the model. 
The reason for the apparent drop of the PD in $B$ is unknown, but we argue that the actual broadband spectrum of an object obtained simultaneously with polarization measurements should be used instead of the model spectrum of the companion star to analyze such subtle effects.
Polarimetric observation in the UV band, in which we expect significant increase in the PD, can unambiguously infer the nature of PA rotation in \onea and, subsequently, shed light on the nature of the UV excess in the binary.

\subsection{On the synchrotron origin of the polarization}
Synchrotron emission has been suggested to contribute to the polarization of \onea in NIR in addition to scattering \citep{Russell2016}. By assuming the scattering component to have polarization with PA $\sim173\degr$ and PD $\sim1\%$ and by subtracting it from the observed $K_{\rm s}$ polarization, authors got the PD of synchrotron emission to be $\sim 1.3\%$. Then, by assuming the jet contribution in $K_{\rm s}$ to be 8--37\%, authors found the jet to be polarized at the level of $\mathrm{PD_{jet}}=$3--18\%. However, we argue that such a separation of polarized components cannot be done unambiguously with the existing data for several reasons.

First, the PA changes significantly within the optical range, meaning that polarization cannot be produced by scattering alone. Therefore, the PA of the scattered component cannot be determined explicitly with the optical data only. In previous section, we estimated the PA of scattered component to be closer to $\sim150\degr$, which, if subtracted from observed $K_{\rm s}$ polarization, would result in much smaller PD of synchrotron emission $\sim$0.8\% with the PA of $\sim$180\degr. We note also that although the observed PA in  $K_{\rm s}$ seems to be perpendicular to the possible direction of the radio jet (see Fig.~\ref{fig:rotation_on_the_sky}), after subtracting the scattering contribution, the PA of remaining synchrotron component differs from the jet axis by about $45\degr$, which cannot be easily explained with the jet model.  Second, there is no direct evidence of the presence of NIR excess in the photometric data in passive quiescence. The apparent increase in the flux in NIR relative to blackbody is well described by the stellar spectrum alone -- the ``bump'' at $K_{\rm s}$ and $H$ is well pronounced in the K5V companion spectrum in Fig.~\ref{fig:spectrum}. Therefore, it is rather difficult to determine the contribution of jet emission to the observed spectrum, especially given the strong variability of the object both within and between states. Adding the jet as a third component in addition to the scattering and UV components to the modeling is hardly justified given the quality and the number of polarimetric observations. Taking all of the above into account, although we do not rule out the possibility that jet contributes to the NIR polarization, we argue that its accurate determination is a challenging task that requires more precise simultaneous multiwavelength polarimetric observations than we have to date.

\begin{figure}
\centering
\includegraphics[width=0.9\linewidth]{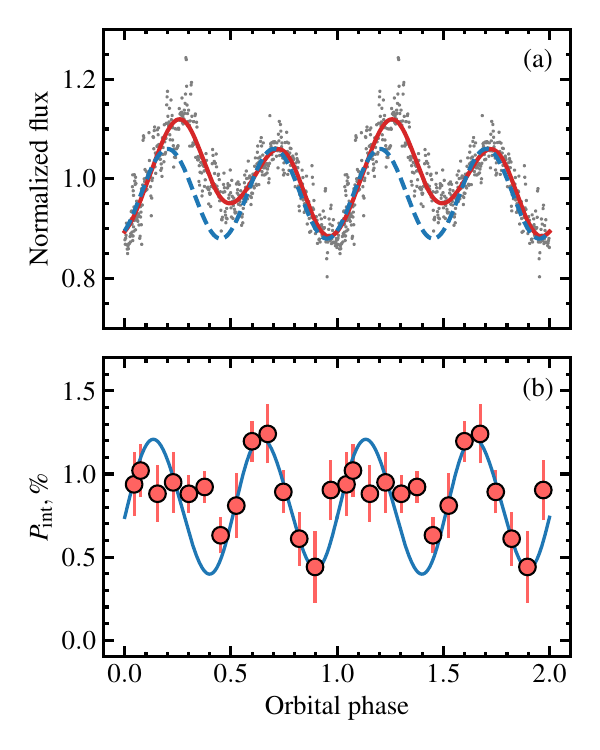}
\caption{Flux and PD variations of \onea. \textit{Panel (a)}: Normalized flux, folded with the orbital phase (gray circles). 
The solid red line corresponds to the best fit of the data with the Fourier series, while the dashed blue line corresponds to the sinusoidal variations, assumed to be produced by the tidally distorted optical companion. 
\textit{Panel (b):}  Intrinsic PD of \onea (red circles) folded with the orbital period together with the model of a scattering cloud on a circular orbit from the appendix of \citealt{Kravtsov2020} (blue line).
\label{fig:model}}
\end{figure}

\subsection{The source of the polarization variations}
\label{sect:variability}

In addition to the changes in the polarization between different spectral states, we also see intra-day variability of the polarization.
Although the $2\%$ orbital modulation of the polarization observed by \cite{Dolan1989} has not been confirmed \citep{Dubus2008}, we see the variations at a level of $0.3\%$ on timescales of hours (see Fig.~\ref{fig:phot_and_pol}). 
To confidently claim that the polarization of \onea is variable with the orbital phase, one would need to cover at least two consecutive orbital periods with high-precision polarimetric observations (which is a rather difficult task given the brightness of the object in quiescence and its short orbital period).\ That said, we can cautiously state that the observed polarization variability is related to the orbital phase rather than having a stochastic nature.
Indeed, the  polarization of \onea, folded with the orbital phase, shows the pattern typical for binary stars -- double-sinusoidal variations with the PD reaching minima in the conjunctions (phases 0 and 0.5) and maxima in the quadratures (phases 0.25 and 0.75). 
This is due to Thomson scattering of the stellar radiation on the matter, which follows the compact object in its orbital motion \citep{BME, Kravtsov2020}, and is similar to what is observed in other BH \citep{Kemp1979, Gliozzi1998, Kravtsov2023} and neutron star \citep{Egonsson1991, Combi2004, Baglio2016} X-ray binaries. 
In Fig.~\ref{fig:model} we see a fairly good agreement of the PD variations with the expected behavior (the blue line shows the model of a scattering cloud on a circular orbit; see the appendix of \citealt{Kravtsov2020}). 
In addition, our first and second observations, performed during two different orbits are consistent with each other (see Fig.~\ref{fig:phot_and_pol}).
We note also that Stokes parameters of \onea, folded with the orbital period, show scatter, similar to one observed in \cite{Dubus2008} – while Stokes $q$ shows more pronounced orbital variability, Stokes $u$ seems to behave more erratically.
The amplitude of this scatter, however, is comparable to the measurement errors, which complicates its interpretation, but one explanation could be the presence of inhomogeneities (clumps) in the scattering matter.

The optical light curve shown in Fig.~\ref{fig:model} has an asymmetric profile -- the peak and the dip at the first half of the orbit are brighter than the other peak and dip. 
The so-called ellipsoidal variations of the flux, produced by the tidally distorted optical star (shown by the dashed blue line at Fig.~\ref{fig:model}), cannot produce such asymmetry; therefore, an additional source of flux variations is needed. 
There are two alternatives: either there is an additional bright component visible only at the first part of the orbit and generating additional flux (e.g., non-circular accretion disk or bright spot in the accretion stream; see, e.g., \citealt{Haswell1993}), or some object blocks the light of the optical companion, reducing the total flux at phases from 0.5 to 1.0 (e.g., dark spots at the surface of the star; e.g., \citealt{Cherepashchuk2019}). 
We see similar asymmetry in the orbital polarization profile of \onea~ -- while around orbital phases from 0.5 to 1.0 the PD shows good agreement with the Thomson scattering model, at the first part of the orbit the PD is consistent with being constant.
The additional bright unpolarized component, visible only at phases from 0.0 to 0.5, could reduce the polarization at these phases and explain both the asymmetric PD and flux variations; therefore, we argue that the bright spot or phase-dependent disk models are more likely than the dark star spot model. 
Another alternative is that the asymmetric PD profile is produced by scattering of stellar radiation off slightly tilted accretion disk.
In this scenario, the illuminated part of the accretion disk is visible to the observer only for half of the orbit \citep[see][]{Kravtsov2023}.
However, scattering alone is not enough to reproduce the $\sim10\%$ increase in flux observed at the first half of the orbit, and the quality of our polarimetric data is not sufficient to discriminate between the complex models.

\section{Conclusions}
\label{sect:conclusions}

We have presented new high-precision phase-resolved optical polarimetric and quasi-simultaneous NIR-to-UV photometric observations of BH X-ray binary \onea in a passive quiescent state. We determined and subtracted the IS polarization, which allowed us to derive the intrinsic polarization of \onea. 
Using that combined with the NIR polarimetric observations from \cite{Russell2016}, we found that the orbit-average intrinsic PA rotates with the wavelength, changing from    
$124\degr$ in the $K_\mathrm{s}$ filter to $177\degr$ in $B$, while the PD remains at the $\sim 1\%$ level throughout this spectral range. Folding our polarimetric observations with the orbital period, we found the significant orbital variability of polarization properties in the $R$ band. 
The shape of the variations, with two minima and two maxima per period, suggests that the polarization is most probably produced by Thomson scattering of the companion star emission off the matter, which follows the BH in its orbital motion (e.g., scattering off the accretion disk or stream).
The lack of variations during the first part of the orbit suggests either that we see a depolarization effect caused by the bright spot visible only at these phases or (less probably) that the scattering material producing the polarization is obscured or tilted relative to the orbital axis.
However, more high-precision polarimetric observations from at least several consecutive orbital periods are needed to draw unambiguous conclusions about the geometry of the scattering medium.

The flat spectrum of the polarization and the presence of the orbital variations suggest that NIR-to-optical polarization has a scattering origin.
However, in that case, the PA is not expected to change with the wavelength as observed. 
We first considered the possibility that stellar radiation scattered off the accretion disk experiences Faraday rotation while traveling toward the observer through the magnetized plasma surrounding the accretion disk. 
The estimated values of the magnetic field and optical depth can be considered consistent with realistic estimates if we take into account our optical points in combination with the NIR points taken almost ten years ago.
On the other hand, the steep trend of our optical points favors a higher Faraday rotation rate, resulting in higher values for the line-of-sight magnetic field. 

We considered the scenario of two polarized components that have different PDs and PAs. 
One component is coming from the stellar scattered light and the other is associated with the additional UV component seen in the spectrum.
The second component may potentially arise from Compton up-scattering of the disk or synchrotron photons in the hot inner flow, similar to what is seen in the quiescent-state low-mass BH X-ray binary MAXI~J1820+070.
The PAs of these components differ by $\sim$40\degr, which translates to either a $50\degr$ or $40\degr$ misalignment between their axes of symmetry.
Future simultaneous polarimetric observations covering the NIR-to-UV range would be extremely helpful in unambiguously determining which of the two models best represents \onea.

\begin{acknowledgements} 
Based on observations made with the Nordic Optical Telescope, owned in collaboration by the University of Turku and Aarhus University, and operated jointly by Aarhus University, the University of Turku, and the University of Oslo, representing Denmark, Finland, and Norway, the University of Iceland and Stockholm University at the Observatorio del Roque de los Muchachos, La Palma, Spain, of the Instituto de Astrofisica de Canarias. 
The DIPol-2 and DIPol-UF polarimeters were built in cooperation between the University of Turku, Finland, and the Leibniz-Institut f\"{u}r Sonnenphysik, Germany.
The Liverpool Telescope is operated on the island of La Palma by Liverpool John Moores University in the Spanish Observatorio del Roque de los Muchachos of the Instituto de Astrofisica de Canarias with financial support from the UK Science and Technology Facilities Council. MAPT acknowledge support from the Agencia Estatal de Investigaci\'on (MCIN/AEI) and the European Regional Development Fund under grant PID2021-124879NB-I00.
This research has been supported by the Finnish Cultural Foundation (VK) and by the Academy of Finland grants 355672 (AV). Nordita is supported in part by NordForsk. 
\end{acknowledgements}

\bibliographystyle{aa}
\bibliography{aanda}

\end{document}